\documentclass[]{spie}
\usepackage[dvips]{graphicx}  

\newcommand{\arcm}{{$^\prime\,$}}
\newcommand{\arcs}{{$^{\prime\prime}\,$}}  
\newcommand{\etal}{{\it et al.}}  

\title{The Deep Lens Survey}

\author{D. Wittman\supit{a}, J.~A. Tyson\supit{a},
I.~P. Dell'Antonio\supit{b}, A.~C. Becker\supit{a},
V.~E. Margoniner\supit{a}, \\ J. Cohen\supit{c}, D. Norman\supit{d},
D. Loomba\supit{e}, G. Squires\supit{f}, G. Wilson\supit{b,f},
C. Stubbs\supit{g}, J. Hennawi\supit{h}, \\ D. Spergel\supit{h},
P. Boeshaar\supit{i}, A. Clocchiatti\supit{j}, M. Hamuy\supit{k}, G. Bernstein\supit{l}, \\ A. Gonzalez\supit{m},
P. Guhathakurta\supit{n}, W. Hu\supit{o}, U. Seljak\supit{h}, D. Zaritsky\supit{p}
\skiplinehalf 
\supit{a}Bell Labs, Lucent Technologies;
\supit{b}Brown University;
\supit{c}Caltech; 
\supit{d}CTIO\\
\supit{e}University of New Mexico;
\supit{f}IPAC/Caltech;
\supit{g}University of Washington\\
\supit{h}Princeton University;
\supit{i}Drew University;
\supit{j}PUCC;
\supit{k}Carnegie Observatory\\
\supit{l}University of Pennsylvania;
\supit{m}University of Florida;
\supit{n}UCO/Lick\\
\supit{o}University of Chicago;
\supit{p}Steward Observatory, University of Arizona
}

\begin{document}
\maketitle
\begin{abstract}

The Deep Lens Survey (DLS) is a deep $BVRz^\prime$ imaging survey of
seven $2^\circ \times 2^\circ$ degree fields, with all data to be made
public.  The primary scientific driver is weak gravitational lensing,
but the survey is also designed to enable a wide array of other
astrophysical investigations.  A unique feature of this survey is the
search for transient phenomena.  We subtract multiple exposures of a
field, detect differences, classify, and release transients on the Web
within about an hour of observation.  Here we summarize the scientific
goals of the DLS, field and filter selection, observing techniques and
current status, data reduction, data products and release, and
transient detections.  Finally, we discuss some lessons which might
apply to future large surveys such as LSST.

\end{abstract}

\section{SCIENCE DRIVERS}

The primary purpose of the DLS is to study the evolution of mass
clustering over cosmic time using weak gravitational lensing.
Secondary goals include studying the relationship between mass and
light through galaxy-galaxy lensing, galaxy and galaxy-quasar
correlation functions, and optical searches for galaxy clusters;
time-domain astrophysics stemming from the transient search and
followup; and enabling a broad array of astrophysics by publicly
releasing a large, uniform dataset of images and catalogs.

The evolution of mass clustering over cosmic time is our primary goal
because to date, most of what we know about the large-scale structure
of the universe comes from the observed anisotropies in the cosmic
microwave background (CMB) at a redshift $z \sim 1100$, and from the
distribution of galaxies at $z \sim 0$.  The concordance model of
cosmology adequately explains both, but to really test the paradigm,
we must examine the evolution between these two extremes.

Weak gravitational lensing is our tool of choice because unlike many
other tools, it is sensitive to all types of matter, luminous or dark,
baryonic or not.  Furthermore, it provides constraints on cosmological
parameters which are complementary to the CMB and to the expansion
history of the universe as probed by Type Ia supernovae.  For example,
certain weak lensing statistics constrain $\Omega_m$ very well while
constraining $\Omega_\Lambda$ only weakly.  The CMB and supernovae
results each constrain different combinations of $\Omega_m$ and
$\Omega_\Lambda$.  The DLS will test the validity of the foundation of
the theoretical model by providing a third, precision measurement
relying on different physics--weak gravitational lensing.


The term ``weak lensing'' actually includes several independent
measurements, all of which take advantage of the fact that shapes of
distant galaxies are distorted when their light rays pass through
intervening mass distributions on their way to us.  The most
straightforward measurement is that of counting mass concentrations
above a certain mass threshold. (Henceforth, we will call these
``clusters'' for simplicity, but it should be understood that weak
lensing is sensitive to all mass concentrations, whether or not they
contain light-emitting galaxies.)  The number of such clusters as a
function of redshift between $z=0$ and $z=1$ is a sensitive function
of $\Omega_m$ and $w$, the dark energy equation of state.

In a complementary measurement, the power spectrum of mass
fluctuations is derived from the statistics of the source shapes
without the need to identify any particular mass concentration.
This power spectrum is expected to grow with cosmic time in a certain
way, so by measuring it at a series of different redshifts, the DLS
will provide sharp constraints on the models.

For visualization, the DLS will produce maps of the mass
distribution at a series of redshifts.  Assembled into a time series,
these maps will show the growth of structure from $z \sim 1$, when the
universe was about half its present age, to the present.


These weak lensing projects imposed the following qualitative
requirements on the data:
\begin{itemize}
\item imaging in 4-5 different filters with a wavelength baseline
sufficient to derive photometric redshifts
\item deep enough to derive photometric redshifts for $\sim$ 50--100 galaxies arcmin$^{-2}$ 
\item subarcsecond resolution, so that shape information about distant
galaxies is retained even after dilution by seeing
\item at least 5 independent, well-separated fields, selected without
regard to already-known structures, to get a handle on cosmic variance
\item in each field, coverage of an area of sky sufficient to include
the largest structures expected
\end{itemize}

Of course deep imaging is useful for a huge variety of other
investigations, so we resolved to retain as much general utilty as
possible while optimizing the survey for lensing.  For example, we
chose widely used filters (described below) rather than adopting a
custom set.

At the same time, we realized that the data would be useful for a
completely different application.  In addition to coadding multiple
exposures of the same field, we could subtract one epoch from another
and find transient phenomena including moving objects in the solar
system, bursting objects such as supernovae and possibly optical
counterparts to gamma-ray bursts, and variable stars and AGN.

Image subtraction, or difference image analysis (DIA), has been used
quite successfully by supernova search groups.  For that
application, however, all other types of transients are nuisances, and
their elimination from the sample is part of the art of the supernova
search.  Ironically, there are other groups using the same 4-meter
telescopes and wide-field imagers, often during the same dark run, to
look for asteroids while discarding supernovae!  We decided to try a
new approach, in which we would do the difference image analysis while
observing and release information about all types of transients on the
Web the same night, so that others could follow up any subsample of
their choosing.  This attempt to do parallel astrophysics is in many
ways a precursor of the much larger survey projects now in the
planning stage which seek to further open the time domain for deep
imaging, and we will discuss some of the lessons learned.

\section{BASIC PARAMETERS}

We chose the Kitt Peak National Observatory (KPNO) and Cerro Tololo
Inter-American Observatory (CTIO) 4-m telescopes for their good
combination of aperture and field of view.  Each telescope is equipped
with an 8k $\times$ 8k Mosaic imager (Muller \etal\ 1998), providing a
35\arcm\ field of view with 0.25\arcs\ pxels and very small gaps
between three-edge-buttable devices.  A further strength is that with
some observing with each of two similar setups, we can provide a check
on systematic errors due to optical distortions, seeing, and the like,
without sacrificing homegeneity of the dataset in terms of depth,
sampling, and so on.

We set our goal for photometric redshift accuracy at 20\% in $1+z$.
This is quite modest in comparison to the state of the art, but we do
not need better accuracy.  A bigger contribution to the error on a
lensing measurement from any one galaxy is shape noise, the noise
stemming from the random unlensed orientations of galaxies.  Thus we
do not need great redshift accuracy on each galaxy, but we do need the
redshifts to be unbiased in the mean.

To keep the total amount of telescope time reasonable, we wanted to
use the absolute minimum number of filters necessary for estimating
photometric redshifts to this accuracy.  Simulations showed that four
was the minimum.  We chose a wavelength baseline from Harris $B$ to Sloan
$z^\prime$, which is the maximum possible without taking a large
efficiency efficiency hit from the prime focus correctors in $U$ or
from a switch to infrared detectors on the other end.  Between these
extremes, we chose Harris $V$ and $R$, where throughput is maximized;
simulations showed that photometric redshift performance did not
depend critically on these choices.

Our depth goal is a 1-$\sigma$ sky noise of 29 mag arcsec$^{-2}$ in
$BVR$ and 28 in $z^\prime$, which required total exposure times of 12
ksec in $BVz^\prime$ and 18 ksec in $R$ according to exposure time
calculators, for a total of 54 ksec exposure on each patch of sky (see
below for discussion).  We divided this total time into 20 exposures
of 600 (900) seconds each in $BVz^\prime$ ($R$).  Twenty exposures was
a compromise between efficiency (given a read time of $\sim 140$
seconds) and the need to take many dithered exposures to obtain good
dark-sky flats.

The metric size of expected mass structures, and the redshift at which
they most effectively lens faint background galaxies, sets the angular
scale for a field.  In order to study the mass clustering on linear
scales (out to $15 h^{-1}$ Mpc at z=0.3), we must image $2^\circ
\times 2^\circ$ regions of the sky. With Mosaic providing a 40\arcm\
field of view after dithering, we assemble each 2$^\circ$ field from a
3$\times$3 array of Mosaic-size subfields.  Thus each field requires a
total exposure time of 9$\times$54 ksec, or 486 ksec.

The number of such fields should be as large as possible to get a
handle on cosmic variance.  To fit the entire program into a
reasonable amount of telescope time, we decided on five fields
observed from scratch plus two others in which we can build on a
previous survey, the NOAO Deep Wide-Field Survey (NDWFS).  The NDWFS
is a $B_WRI$ survey of two 9 deg$^2$ fields with partial $JHK$
coverage as well.  We decided to locate a DLS field within the
infrared-covered area of each of these fields, and simply add
additional $R$ imaging (10 exposures of 900 seconds on each patch of
sky) to bring it to the same $R$ depth as the remainder of the DLS.
Thus the total exposure time for piggybacking on the NDWFS fields is
9ksec times 18 subfields, or 162 ksec.  The total exposure time for
the entire survey is then 2592 ksec.

The number of nights is then determined by the observing efficiency.
The read time of the Mosaic was $\sim$ 140 seconds at the start of the
survey, implying about 82\% efficiency if the observers are able to
focus and take standard star images during twilight.  The read time
was slated to decrease significantly after an upgrade from 8 to
16-amplifier readout, and so we felt 82\% efficiency was a reasonable
goal for the entire survey (see below for discussion).  Thus we needed
3 Msec of time, which is 86 10-hour nights.  We split this between
KPNO and CTIO by assigning two new and one NDWFS field to KPNO and
three new and one NDWFS field to CTIO.

\section{FIELD SELECTION}

The first consideration in field selection was availability of deep
spectroscopy for calibrating photometric redshifts in at least one
field.  After choosing that one field, the gross location of the other
fields fell into place through traditional considerations such as
avoiding the plane of the Galaxy and spacing the fields to share
nights smoothly.  The precise location was then chosen to avoid bright
stars as much as possible.  Fields were not chosen with regard
to known structures such as clusters.

We chose the Caltech Faint Galaxy Redshift Survey (Cohen \etal\ 1999)
for our deep spectroscopy, so we centered our first field (F1) in the
north at 00:53:25.3 +12:33:55 (all coordinates are J2000).  The second
field (F2) must then be $\sim$5 hours earlier or later to be able to
share nights with this field.  Earlier is impractical because it would
require summer observing during the monsoon season at Kitt Peak.  A
later field would have to be further north to avoid the plane of the
Galaxy.  We chose 09:18:00 +30:00:00 because it had few bright stars
and low extinction.

In the south, we wanted our fields complementary in RA to those in the
north, so that we would not have simultaneous observing runs in the
same month.  It was difficult to find three southern fields for the
5--15 h range evenly spaced in RA, while avoiding the galaxy.  The
middle of the three (F4) was fixed at 10:52 -05:00 because redshifts
would be available from the 2dF Galaxy Redshift Survey (2dFGRS,
Colless \etal\ 2001).  The 2dFGRS covers vastly more area, but this
happens to be a region of low extinction.  (The 2dFGRS goes to $z\sim
0.3$, and thus is of much less importance than the CFGRS in
calibrating our photometric redshifts, which we want to push to $z\sim
2$.  However, we felt that any additional spectroscopy would help.)
The late field (F5) was chosen to be very near the ecliptic to help
the hunt for Kuiper belt objects (KBOs); 13:55 -10:00 was chosen for
low extinction, and partial overlap with the Las Campanas Distant
Cluster Survey (LCDCS; Gonzalez \etal\ 2001).  The early field (F3)
was the most difficult to place because it was up against the Galaxy.
We chose 05:20 -49:00, which at a Galactic latitude of -35 is the
closest field to the plane.

Note that proximity to the plane is really shorthand for two different
things: Galactic extinction, which we extracted from the maps of
Schlegel et al, and bright stars, for which we consulted star atlases.
Stars in general are actually helpful for lensing, because they reveal
the point-spread function (PSF), but they become useless when
saturated (at around 18.5$^m$ in most of the $R$ exposures), and in
fact become destructive when they are bright enough to prohibit
detection and evaluation of galaxies projected nearby (a tenth
magnitude star causes a significant amount of lost area).  These two
aspects are not perfectly correlated, so while F3 is closest to the
plane and has the most bright stars, it does not have the greatest
extinction.

The NDWFS (F6 and F7) is still completing their observations, so we
decided to postpone exact location of our fields within theirs until
late in our survey, when we should know where the imaging quality and
wavelength coverage are best.

\begin{figure}
\begin{center}
\begin{tabular}{|c|c|c|c|c|} 
\hline 
Field & RA & DEC & l,b & E(B-V) \\ \hline 
F1 & 00:53:25.3  &    +12:33:55   &    125,-50 & 0.06 \\
F2 & 09:18:00    &    +30:00:00   &    197, 44 & 0.02 \\
F3 & 05:20:00    &    -49:00:00   &    255,-35 & 0.02 \\
F4 & 10:52:00    &    -05:00:00   &    257,47  & 0.025 \\
F5 & 13:55:00    &    -10:00:00   &    328,49  & 0.05 \\
F6 & 02:10:00    &    -04:30:00   &    166,-61 & 0.04 \\
F7 & 14:32:06    &    +34:16:48   &    57,63   & 0.015 \\ \hline 
\end{tabular} 
\end{center} 
\caption{Field information.  Coordinates are J2000, at the field
center. F6 and F7 are subsections of the NOAO Deep Wide- Field Survey,
and their exact positions within the larger NDWFS fields will not be
set exactly unto; the image quality and wavelength coverage variations
of the NDWFS are known.  Extinctions are from the maps of Schlegel et
al and represent rough averages over large areas which vary
somewhat. }
\end{figure}

\section{OBSERVING STRATEGY}

Subarcsecond resolution is common for individual exposures on these
telescopes, but achieving this resolution in the final product of a
large survey is another matter, as a large range of seeing conditions
is to be expected.  We address this problem by observing in $R$ when
the seeing conditions are 0.9\arcs\ or better, and switching to $B$,
$V$, or $z^\prime$ when the seeing becomes worse than 0.9\arcs.  Thus
the final $R$ images are guaranteed to be 0.9\arcs\ or better.  We
will then make the weak lensing shape measurements in $R$ and use the
other filters only for color information.

This seeing dependence is the main element in deciding what to observe
when at the telescope.  The next most important consideration is
moon---when there is significant moon illumination, we observe in
$z^\prime$, which is the least sensitive to such illumination.  Next,
there is observing efficiency.  We can offset, but do not wish to
slew while reading out, so the most efficient tactic is to take as
many consecutive dithered exposures as possible on a given subfield.
Dithers are up to $\pm$200\arcs, or half the width of a 2k$\times$4k
device, to insure good superflats.  Although we could change filters
rapidly while reading out, the filter is mostly dictated by
atmospheric conditions as described above.  Thus we settled on a
pattern of five dithered exposures in any given subfield/filter
combination before slewing to another subfield (and possible changing
filters if conditions dictate).  This still leaves several degrees of
freedom, which we grant to the transient search.

We wish to sample a range of timescales, from minutes to years.  As
described above, we spend of order one hour on a given subfield/filter
combination.  This timescale is perfect for discovering inner solar
system objects (in fact it is used by the supernova searches to
eliminate asteroids).  It is also largely unexplored for more exotic
bursting objects; supernova searches do take exposures on a given
field one hour apart, but with only two data points, anything that
does not behave as expected is simply thrown out.

The next priority is to capture the one-month timescale, which has a
known population of interesting transients: supernovae.  We schedule
runs a month apart for sensitivity to supernovae, and make every
attempt to observe subfields and filters which were observed the
previous month.  At KPNO, we have two runs each year separated by a
month, and at CTIO, three runs each separated by a month, barring
scheduling difficulties.

Finally, we also have day (suitable for searches for rans-neptunian
objects and possible optical afterglows of gamma-ray bursts) and year
(suitable for supernovae and AGN) timescales.  We have runs of 4--6
nights each, so in the latter half of a run the priority is to revisit
a subfield/filter combination done in the first half of the run.  And
one-year timescales are involved because we typically do not finish all
twenty exposures in a given subfield/filter combination in a single
observing season.

\section{DATA REDUCTION}

The following steps are performed at the end of a run using IRAF's
{\it mscred} package:
\begin{itemize}

\item Astrometric solution: each image comes is matched to the USNO-A2
star catalog using the {\it msccmatch} task).  After this step, the
uncertainty in the global RA,DEC coordinate system relative to the
USNO-A2 is less than 0.01\arcs.  The rms error in each object's
position is $\sim$0.3\arcs, which is mostly limited by the A-2
catalog's internal accuracy.
\item Crosstalk correction: We use the NOAO-determined 
corrections to mitigate the crosstalk which occurs as multiple
amplifiers are read in parallel.  This step and the overscan, zero,
and flattening are done with the {\it ccdproc} task.
\item Overscan subtraction 
\item Zero subtraction 
\item Flatfield pupil removal: in the KPNO $z^\prime$ data, a ghost
image of the pupil appears as an additive effect in the dome flats.
We use the {\it rmpupil} task to remove it.
\item Dome-flat correction
\item Sky-flat correction: the sky flat is made from a stack of all
the deep dome-flattened images in a given filter from the entire run.
\item Pupil removal (KPNO $z^\prime$ data only) 
\item Defringing ($z^\prime$ data only; {\it rmfringe} task) 
\end{itemize}
These are performed in the traditional way, by manually running the
tasks and inspecting the results.  We feel that our mission is
sufficiently small that every exposure should be inspected.  At the
same time, the tools to make an effective pipeline from existing IRAF
tasks were not in place when we started the survey.  With the
emergence of tools like PyRAF, we leave open the possibility that a
final re-reduction of all the inspected data may be done with an
automated pipeline.

\begin{figure}
\centerline{\resizebox{3in}{!}{\includegraphics{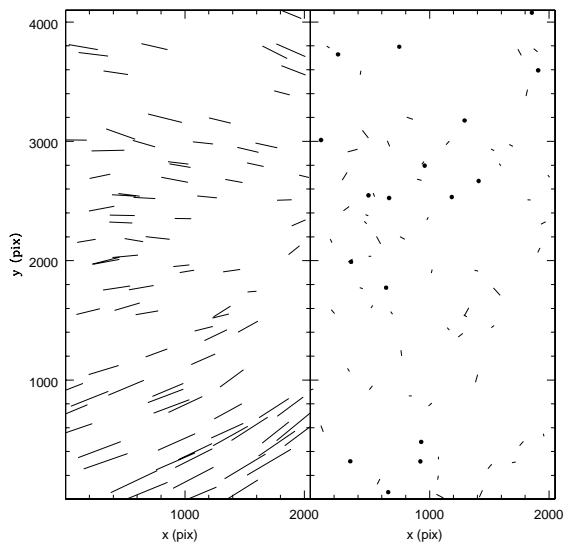}}
}
\caption{Point-spread function correction in one 2k$\times$4k CCD.
Shapes of stars, which as point sources should be perfectly round, are
represented as sticks encoding ellipticity and position angle.  Left
panel: raw data with spatially varying PSF ellipticities up to 10\%.
Right panel: after convolution with a spatially varying asymmetric
kernel, ellipticities are vastly reduced (stars with $\epsilon<0.5\%$
are shown as dots), and the residuals are not spatially correlated as a
lensing signal would be.}
\label{fig-circ}
\end{figure}

Conceptually, the next step is the lensing-specific step of making the
PSF round.  Because this is specific to lensing, this is where our
custom software takes over; we view the flatfielded data as ``raw''
data for our lensing pipeline.  Figure~\ref{fig-circ} illustrates the
problem: an anisotropic PSF mimics a gravitational lensing effect in
many ways, with shapes of objects correlated at least locally.  PSFs
on real telescopes tend to be $\sim 1-10\%$ elliptical, which is much
larger than the lensing signal we are looking for, and thus must be
eradicated as much as possible.  This is done by convolving the image
with a kernel with ellipticity components opposite to that of the
PSF (Fischer \& Tyson 1997). Because the PSF is 
position-dependent, the kernel must be also.

However, this correction cannot be applied immediately, as each image
must be remapped onto a common coordinate system (optical distortions
reach 30\arcs\ at the corners of the Mosaic), and this remapping
itself changes the PSF shape.  Therefore the software must catalog
each image; set up a coordinate system for the final image; match
catalogs and use the matches to determine the transformations to the
final coordinate system, rejecting outliers; select stars; compute
what the PSF shape will be after remapping (to avoid the computational
expense of writing out intermediate images); fit for the spatial
variation of the stellar shapes, rejecting outliers; determine the
photometric scaling between images from the matched catalogs; and
combine the images using all these transformations.  Most of these
tasks are implemented as standalone programs written in C, and they
are executed in the proper order by a Perl script, {\tt dlsmake},
which also checks the return values, and often the output data, of the
tasks for potential errors.  Human intervention is required only for
inspection of the star selection, which sometimes goes awry. There is
a graphical user interface which displays the magnitude-size plane and
allows the user to adjust the stellar locus by dragging a box over it.

Precise registration is crucial, as a small registration error can
mimic a shear.  The rms departure of object centroids from the
coordinate fit prediction is $\sim$ 0.1 pixel, or 0.025\arcs.  To
nullify any PSF shape systematics which might have been introduced by
this, we select a new sample of stars from the coadded image, and
convolve it with the appropriate spatially varying kernel.

We make extensive use of existing tools.  For example, we use
Sextractor (Bertin \& Arnouts 1996) to make quick catalogs of each
input image for the pipeline, even though we have not settled on it
for the science-grade cataloging of the final images.  At the same
time, we ask Sextractor to produce a sky-subtracted version of each
input image.  We use these sky-subtracted versions in the combine, so
that chip-to-chip sky variations do not produce artifacts in the final
image.  These are the only intermediate images written to disk; the
resampled and convolved versions reside only temporarily in buffers in
the combine program, {\tt dlscombine}.

Throughout, attention is paid to masking.  In addition to
straightforward bad-pixel masking, we must deal with plumes of
scattered light from very bright stars off the edge of the field in
some individual exposures.  We use ds9 to draw regions around these
and save a link to the region file in the image header.  {\tt dlsmake}
then makes a custom bad-pixel mask on the fly, but applies it only
where relevant.  For example, the region is still examined for sources
which will help determine the coordinate transformation or the spatial
variation of the PSF; otherwise an image could not be processed at all
if a substantial part of it must
be masked.


\section{TRANSIENTS}

In a separate pipeline, we use difference image analysis to subtract
off the temporally constant parts of our images and reveal the
variability.  This pipeline is now managed using the Opus data
processing environment (Miller \& Rose 2001 and references therein),
and runs while observing.  Thus it must use afternoon dome flats
instead of sky flats.  The difference in image quality is not great,
but it is the main reason that these reductions are carried out
separately from the deep imaging reductions.

We use a modified version of the Alard (2000) algorithm 
to register both the height and shape of the
individual point-spread functions.  We difference the individual
dithers for short timescale events down to $\sim 23$rd magnitude, and
co-add the images for comparison to a previous co-add for sensitivity
down to $\sim 24.5$th mag.  Detection and some winnowing of spurious
events (cosmic rays, bleeds from bright stars) is done automatically,
but the observer is required to do the final classification.  The main
motivation here is that events are posted to the web as soon as they
are approved, so human-caliber quality assurance is more important
than maintaining strict objectivity, even if that makes efficiency
calculations difficult.  The website ({\tt
http://dls/bell-labs.com/transients.html}) lists positions, dates,
magnitudes, classifications, and has time series images similar to the
figures shown here.  We find several hundred transients per run, so
entries are color-coded according to classification.  For each moving
object, the angular velocity is listed, and an ephemeris is
available.  For stationary objects, a finding chart is available.

\begin{figure}
\centerline{\resizebox{3in}{!}{\includegraphics{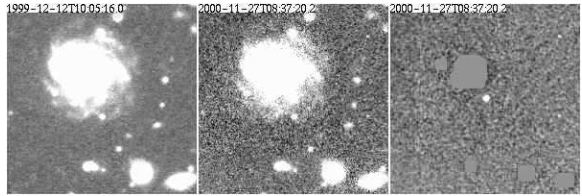}}}

\caption{Input and output of the difference imaging
technique.  The left-most panel shows the template, or comparison
image, at a previous epoch.  The supernova contained in the second
epoch image, central panel, is not evident.  However, the difference
of the two image, right-most panel, subtracts the entire galaxy off,
revealing this transient event.}

\label{fig-trans219}
\end{figure}

Figure \ref{fig-trans219} shows how well the image subtraction works,
revealing a supernova otherwise enmeshed in the isophotes of an
extended host galaxy.  Over the couse of the survey, we have detected
nearly 100 supernova candidates, and a similar number of transients
close enough to the nuclei of galaxies to potentially be AGNs.  Over
the last year we have made a concerted effort to reacquire supernova
candidates on subsequent nights (possibly in other filters) to meet
the requirements of the International Astronomical Union for
designating official supernova.  Thus, 12 out of 14 of our designated
supernova have come in the past year.  We hope to coordinate and
distribute upcoming supernovae to next-generation supernova
projects.

\begin{figure}
\centerline{\resizebox{3in}{!}{\includegraphics{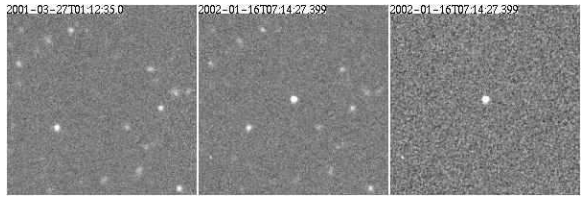}}}
\caption{Transient event with no apparent host.  The temporal baseline
between the template and comparison image was approximately 10 months.
Photometric follow-up indicated the object was very blue ($V - R \sim
0.08$).  Spectroscopy from the Las Campanas Observatory 6.5-m Baade
telescope was inconclusive, but showed a blue featureless continuum.
This information combined with a lack of temporal variability over
several days suggests this object was possibly a Type II SN.}
\label{fig-trans139}
\end{figure}

We have had suceess obtaining follow-up observations for the
interesting cases where there is no apparent host for the transient.
In the case of the transient in Figure \ref{fig-trans139}, from Jan
2002, we were able to obtain both photometric and spectroscopic
follow-up observations after disseminating information on this event
over the Variable Star NETwork
\footnote{http://vsnet.kusastro.kyoto-u.ac.jp/vsnet/}, and expect to
further our response by using the Astronomer's Telegram
\footnote{http://atel.caltech.edu/}.

\begin{figure}
\centerline{\resizebox{3in}{!}{\includegraphics{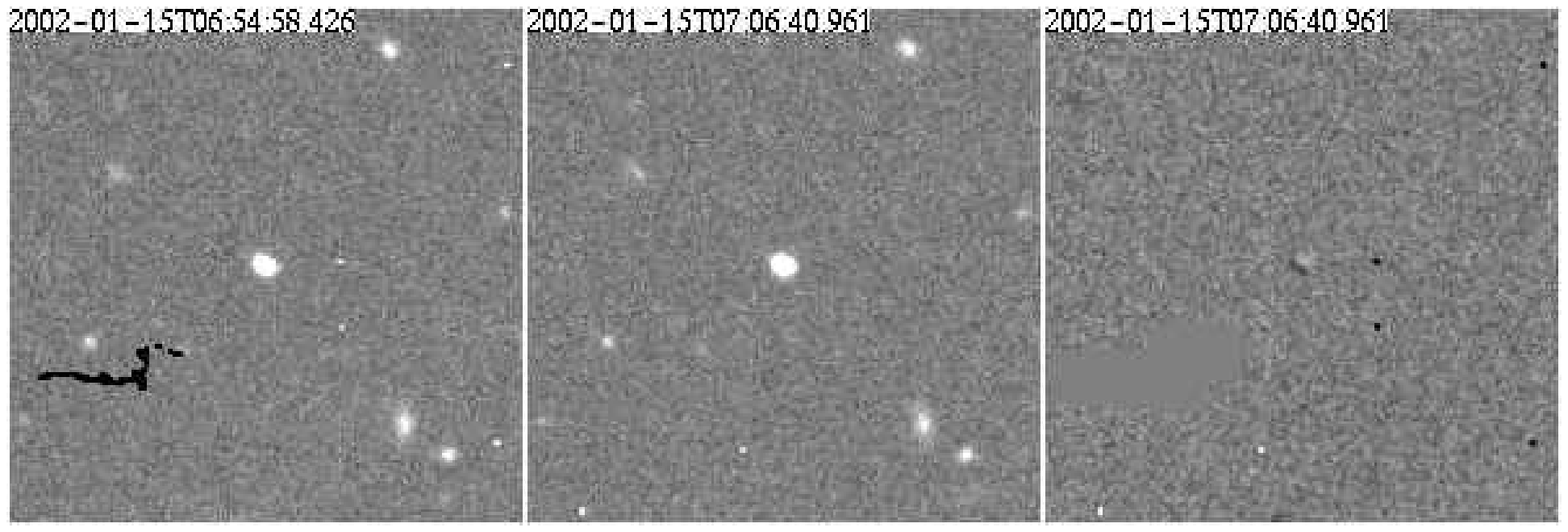}}}
\centerline{\resizebox{3in}{!}{\includegraphics{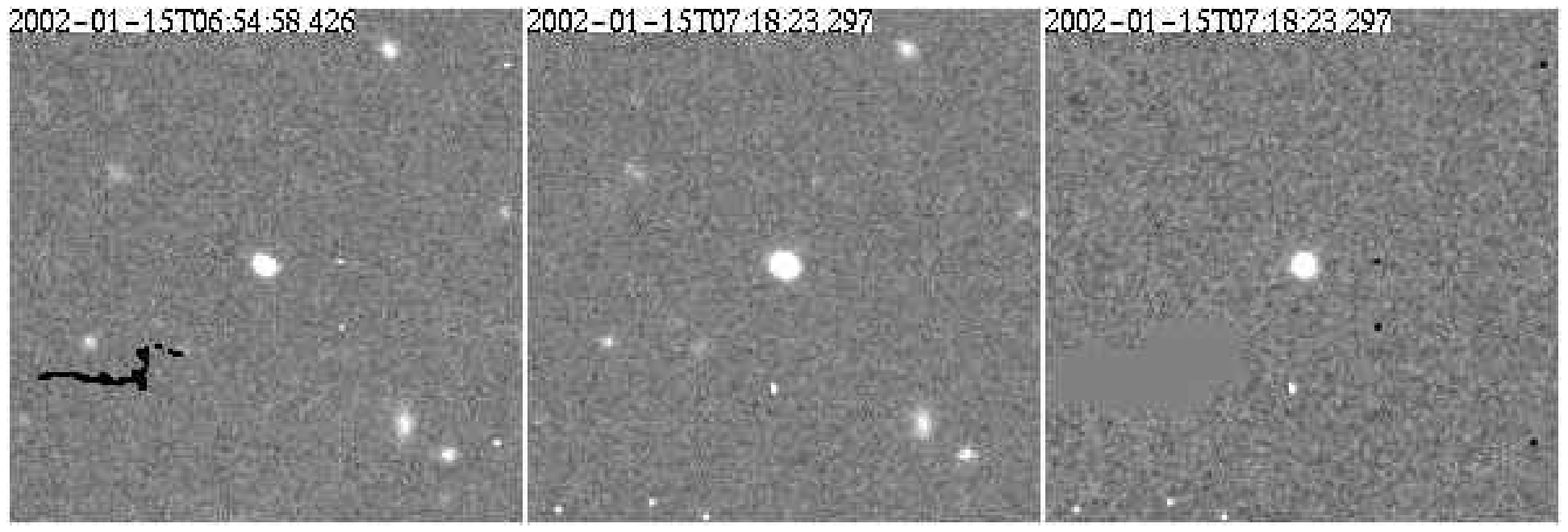}}}
\centerline{\resizebox{3in}{!}{\includegraphics{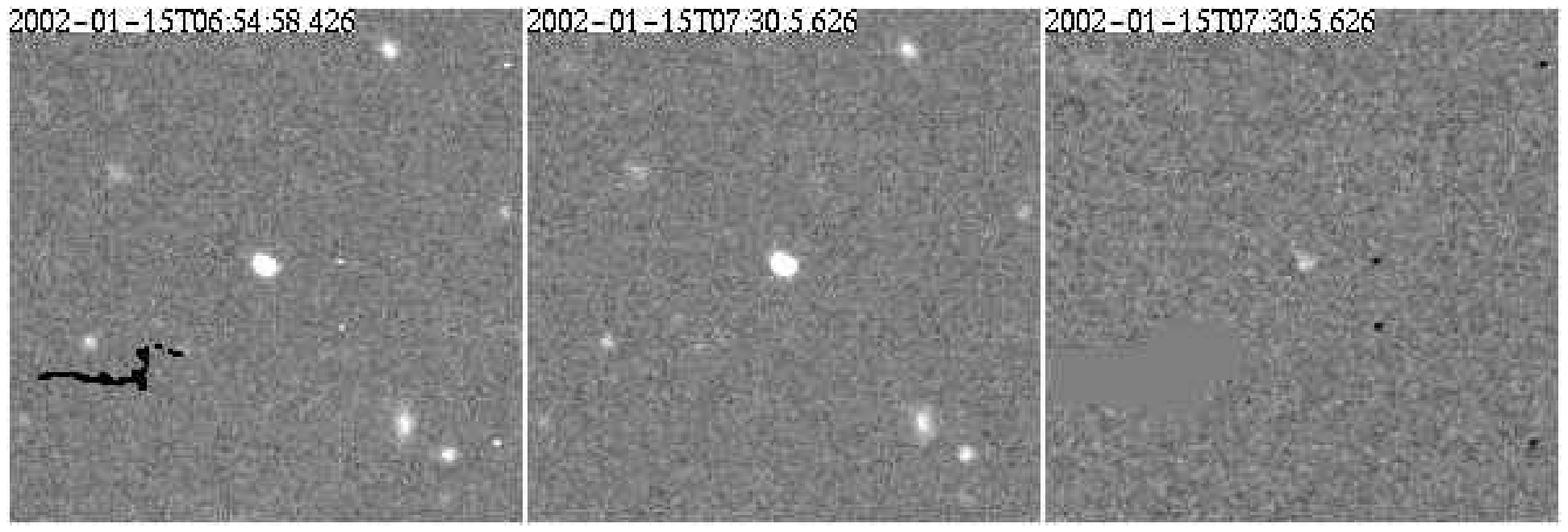}}}
\caption{A bright bursting event apparently in a faint host galaxy.
Images are separated by 600 seconds.  Subsequent spectroscopy showed
that the line of sight to the unsuspecting galaxy contained a flare
star. }
\label{fig-trans337}
\end{figure}

In the most interesting of the classes of transients, rapidly varying
phenomena, we have a very small crop of candidates, none of which have
proved to be extragalactic.  Figure ~\ref{fig-trans337} shows an
exaple, in which a burst was detected over what appears to be a
background galaxy.  This burst appeared in the third of five
consecutive 600-second exposures, then faded rapidly.  We immediately
obtained follow-up spectroscopy on the Las Campanas Observatory 6.5-m
Baade telescope, showing the spectral signatures of an M4 dwarf with
H-$\alpha$ in emission---a foreground Galactic flare star---on top of
a faint blue background galaxy.  Exotic-looking events with low event
rates are not necessarily astrophysically interesting!

We also have several thousand temporal sequences of solar system
objects down to $23$rd magnitude.  
Preliminary analysis of this sample confirms the finding by Izevic
\etal\ (2001) of a break in the power-law slope of the size
distribution.  The Minor Planet Center has awarded preliminary
designations to approximately 200 of these objects.

\section{CURRENT STATUS} 

Observing started in the fall of 1999.  After three observing seasons,
we have completed 7 of the 63 subfields.  The remaining coverage is
fairly random due to the twin constraints of switching to $R$ in good
seeing and maximizing the transient productivity.  Thus most subfields
are not yet useful scientifically, because the coverage is too shallow
and/or limited to a few filters.  In the NDWFS fields, the $R$ data
upon which we intended to piggyback is unsuitable due to poor seeing
(lensing is not one of the drivers of that survey), so those fields
are very far behind.  The full status of each subfield and filter is
best visualized at {\tt http://dls.bell-labs.com/progress.html}, which is
updated after each run.  We released deep images and catalogs of the
first two completed subfields in August 2001, and will release five
more in August/September 2002.  Refer to the DLS website
{\tt http://dls.bell-labs.com} for the latest details.

\begin{figure}
\centerline{\resizebox{5in}{!}{\includegraphics{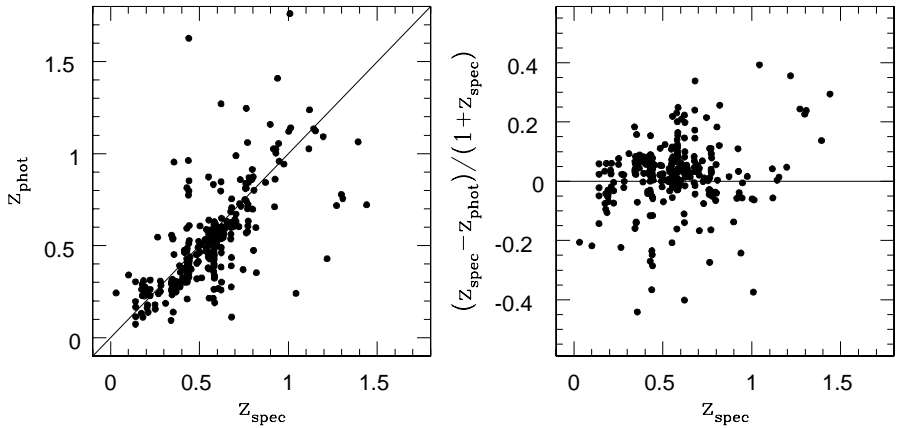}}}
\caption{Upper panel: photometric versus spectroscopic
redshift in a subfield of F1 including hundreds of galaxies from the
CFGRS.  Lower panel: residuals versus spectroscopic
redshift.}
\label{fig-zphot}
\end{figure}

We found that the night sky was significantly brighter in $R$ and
$z^\prime$ than assumed by the exposure time calculators, because we
started observing near solar maximum while the calculators had been
calibrated closer to solar minimum.  We are maintaining the planned
exposure times while assessing the impact on depth.  It appears that
we will still reach our goal of 50--100 galaxies arcmin$^{-2}$ with
photometric redshifts and shape measurements; we are getting around 60
in the few subfields which are up to full depth.  Of course, there
is a tradeoff between density of sources and accuracy of redshifts,
and that is the subject of continuing refinement.  Continuing
refinement should not obscure the overall good performance of the
photometric redshifts. Figure~\ref{fig-zphot} shows that the rms error
is less than 20\% in $1+z$, with very little bias ($<$1\%), using the
template-fitting method plus a magnitude prior.

The first lensing results, focusing on specific subfields rather than
large areas, are coming out very soon.  The first is a paper on the
discovery and tomography of a cluster at $z=0.68$ (Wittman \etal\
2002).  This is only the second cluster discovered via weak
lensing, and its redshift is more than twice that of the first such
cluster.  This bodes well for cluster searches
over the entire DLS area, and the tomography confirms that the
photometric redshifts are useful.  Several other clusters have been
discovered; two are already-known Abell clusters.

Careful attention is paid to quality assurance in the cataloging of
the science-grade data, but we have found that there is no substitute
for trying to do science with it.  Quality assurance as a
self-contained task is bound to fail, no matter how assiduously
carried out.  To that end, work is starting on many different ways of
stressing the data, including cosmic shear, angular correlation
functions, searches for cool stars, quasar-mass correlations, and
using the photometric redshifts to assess intrinsic ellipticity
correlations.  We have already gone through several iterations of
refining our science-grade cataloging procedure.  That procedure is
still preliminary, and the final version will be described in future
papers.

\section{LESSONS FOR FUTURE LARGE SURVEYS}

We have learned many lessons, but only a few are applicable to future
large surveys with dedicated instruments such as LSST (Tyson \etal\
2002; Angel \etal\ 2001; see also these proceedings).

First, take a long-term view.  This is obvious, but it is difficult to
integrate into every aspect of the survey and every tool used to
design the survey.  A good example is the exposure time calculator
having been calibrated at solar minimum.  To catch this kind of error,
a long-term mentality must be very deeply ingrained.

Second, every conceivable data quality check must be done
automatically.  Our data volume was small enough to inspect manually,
but we still encountered a variety of failure modes richer than our
imaginations.  Each instrument has its own peculiarities, so it is
vital to have some early data {\it from the instrument} to help refine
the software.  Simulated data will not do.

Another aspect which simply will not scale is our manual
classification of transients.  Automatic classification is a problem
which has not been tackled because it hasn't been necessary, and
therefore we really don't know how difficult it will be.  Thus, work
should be started now, well in advance of any flood of data, and we
are planning to do so in the next year.  The example of the flare star
in front of a faint galaxy shows that exploring low-event-rate physics
is complicated by all kinds of other low-event-rate processes.

Finally, we have been very successful at finding transients, but
followup has been extremely difficult to coordinate.  The problem
scales roughly as the factorial of the number of different
observatories involved.  Furthermore, it is easy to detect transients
which are too faint to reach with spectroscopy.  Therefore,
time-domain experiments like LSST must design in as much self-followup
as possible by revisiting fields at a useful cadence.  One thing that
would help greatly is obtaining colors right away, which unfortunately
wasn't possible given the design of the DLS.  Simply finding
transients turned out to be the easy part; the challenge is in
followup and in quickly measuring features which enable effective
followup.

\section{ACKNOWLEDGEMENTS}

We thank the KPNO and CTIO staffs for their invaluable assistance.
NOAO is operated by the Association of Universities for Research in
Astronomy (AURA), Inc. under cooperative agreement with the National
Science Foundation.


\end{document}